\documentstyle[preprint,aps,a4,12pt]{revtex}
\begin{document}
\draft
\hfill\vbox{\baselineskip14pt
            \hbox{\bf KEK-TH-510}
            \hbox{KEK Preprint xx-xx}
            \hbox{\today}}
\baselineskip20pt
\vskip 0.2cm 
\begin{center}
{\Large\bf The Relevant Operators for the generalized time-dependent 
m-photon Jaynes-Cummings Hamiltonian}
\end{center} 
\vskip 0.2cm 
\begin{center}
\large S.~Alam\footnote{Permanent address: Department of Physics, University
of Peshawar, Peshawar, NWFP, Pakistan.}, C.~Bentley\footnote{Prairie 
View Texas A \& M University, Texas, USA.}
\end{center}
\begin{center}
{\it Theory Group, KEK, Tsukuba, Ibaraki 305, Japan }
\end{center}
\vskip 0.2cm 
\begin{center} 
\large Abstract
\end{center}
\begin{center}
\begin{minipage}{14cm}
\baselineskip=18pt
\noindent
%%%%%%%%%%%%%%%%%%%%%%%%%%%%%%%%%%%%%%%%%%%%%%%%%%%%%%%%%%%%%
% This is the abstract
%\begin{abstract}
The m-photon Jaynes-Cummings Hamiltonian is a natural generalization
of the much studied Jaynes-Cummings Hamiltonian. In this short note
we give the relevant operators for the time-dependent generalized 
m-photon Jaynes-Cummings Hamiltonian. The dynamical equations for these
operators are also given. These operators are needed and indeed
are the basic building blocks for performing calculations in the
context of the Maximum Entropy Formalism.
%\end{abstract}
\end{minipage}
\end{center}
\vfill
\baselineskip=20pt
\normalsize
\newpage
\setcounter{page}{2}
%------------------------------------------------------------------------ 
\section{Introduction}
%\twocolumn
In a series of papers \cite{ali91,gru93,gru95}, the generalized
time-dependent Jaynes-Cummings Hamiltonian in the context of Maximum
Entropy Principle [MEP] and group theory based methods \cite{pran90} 
was studied. In particular, in \cite{ali91} the MEP formalism was 
used to solve time-dependent N-level systems. A set of generalized 
Bloch equations, in terms of relevant 
operators was obtained and as an example the $N=2$ case was solved.
It was thus demonstrated in \cite{ali91} that the dynamics and 
thermodynamics of a two-level system coupled to a classical field 
can be fully described in the framework of MEP and group theory
based methods. 
Further in \cite{gru93} a time-dependent generalization of the JCM
was studied and by showing that the initial conditions of the operators
are determined by the MEP density matrix the authors were able to
demonstrate that inclusion of temperature turns the problem into a
thermodynamical one. An exact solution was also presented in the
time independent case. Finally in \cite{gru95} more detailed analysis
of the three set of relevant operators was given. These set of
operators are related to each other by isomorphisms which allowed
the authors to consider the case of mixed initial conditions.

	As is well-known when we consider a quantum two-level 
system interacting with a single mode of quantized field one 
is led to the familiar Jaynes-Cummings Hamiltonian [JCH] 
\cite{jay63} provided one is interested only in the difference
of the population of the two levels. The JCH has been extensively 
used as a model Hamiltonian in fields such as 
quantum optics, nuclear magnetic resonance, and quantum 
electronics. A interesting study
of the JCH is the periodic spontaneous collapse and revival 
due to quantum granularity of the field \cite{mey90}.
In the rotating wave approximation [RWA], the JCH
becomes solvable and it has been broadly used in the last years
\cite{eber80,rr89,hong92,ho92,mill92,pran92}.

	The mean values of the field's population, correlation
functions and $n$th-order coherence functions are of interest
and useful in several applications.
The MEP formalism allows us to describe a Hamiltonian system
in terms of those, and only those, quantum operators 
{\it relevant} to the problem at hand. Thus, this formalism 
is suitable to study the Hamiltonian given 
in \cite{gru93,gru95}. In \cite{gru93,gru95}
the {\em population of each level and not their difference} 
is considered therefore the resulting Hamiltonian is 
called a {\em generalized} time-dependent JCH.

	Our aim is to consider the m-photon generalized
time-dependent JCH. The m-photon JCH describes the interaction
of a two-level system with a single quantized mode of
electromagnetic field via m photon emission and absorption
processes between the two-levels. A couple of forms of
the m-photon JCH have been suggested \cite{buck81,suk84}
one of them being intensity dependent.
This simply means that the coupling of this m-photon JCH
is proportional to the square root of the number operator
for the photons. As the validity of this effective Hamiltonian
may be questionable under certain circumstances\cite{bar91}, we 
do not consider it in this paper. We consider the time-dependent
generalized version of the simpler m-photon JCH in this paper. 
The question arises does at least one set of relevant operators
exist for the m-photon JCM? The aims of this short note is to give
such a set and to give the evolution equations for it.
The layout of this paper is as follows. In the next section we
recall some well-known results of the group theory based MEP
formalism. In section two we give the relevant operators and
the evolution equations for their expectation values i.e.
we use generalized Ehrenfest theorem to obtain Bloch equations.   
%%%%%%%%%%%%%%%%%%%%%%%%%%%%%%%%%%%%%%%%%%%%%%%%%%%%%%%%%%%
\section{Summary of the MEP Formalism}
%\twocolumn
 It is instructive to summarize the principal concepts of the 
MEP \cite{ali91,gru93,gru95,alh77,pro89}.
Given the expectation values $<\! \hat O_{j} \!>$ of the operators
$\hat O_{j}$, the statistical operator $\hat \rho(t)$ is defined by
%$\hat\rho_{(t)}$
\begin{equation}
\hat \rho(t)= 
\exp\left(-\lambda_{0}\hat I-\sum_{j
=1}^{L}\lambda_{j}\hat O_{j}\right),\label{F1}
\end{equation}
% \\[0.3ex]
where $L$ is a natural number or infinity, and the $L+1$ Lagrange
multipliers $\lambda_{j}$, are 
determined to fulfill the set of constraints
\begin {equation}
<\! \hat O_{j} \!> = {\rm Tr} \;[\: \hat \rho(t) \;  \hat O_{j} \:] \; ,
\hspace{1.0cm} \mbox{j = 0, 1, \ldots, L} \; ,\label{F2} 
\end{equation}
($\hat O_{0}=\hat I$ is the identity operator) and the normalization
in order to maximize the entropy, defined (in units of the Boltzmann
constant) by
\begin {equation}
S(\hat \rho) =-{\rm Tr}\;[\: \hat \rho  \;  \ln \hat \rho \:] \; .\label{F3}
\end{equation}
Eq.~\ref{F1} is a generalization of the more familiar
density operator. For e.g. in open system, where we
have Grand Canonical Ensemble there are two Lagrange 
multipliers, $\beta=\frac{1}{k_B T}$ and $\mu$  are present, 
and we write the density operator as \cite{rei87} 
\begin{equation}
\hat \rho(t)=
\exp\left(\beta\Omega(T,V,\mu)-\beta \hat{H}+\beta\mu \hat{N}\right),
\label{F1x}
\end{equation}
As is well-known the dynamics are governed by the time evolution 
of the  statistical operator.
The time evolution of the statistical operator is given by
\begin {equation}
i\hbar \frac{d \hat \rho}{dt}=[\: \hat H(t), \hat \rho(t) \:] 
\; .\label{F4}
\end{equation}
The essence of the MEP formalism in conjunction with the
group theory method is to find the relevant operators 
entering Eq.~\ref{F1}) so as to guarantee not only 
that $S$ is maximum, but also is a constant of
motion.~Introducing the natural logarithm of Eq.~\ref{F1} 
into Eq.~\ref{F4} it can be easily verified that the 
{\bf relevant operators}
are  those that close a semi-Lie algebra under 
commutation with the Hamiltonian  $\hat H$, i.e.
\begin {equation}
[\: \hat H(t),\hat O_{j} \:]=
i\hbar\sum_{i=0}^{L}g_{ij}(t)\hat O_{i}\; .
\label{F5}
\end{equation}
Thus the relevant operators may be defined as those satisfying
the above equation. Eq.~\ref{F5} defines an $L \times L$ 
matrix $G$ and constitutes the central requirement to be 
fulfilled by the operators entering in the density
matrix. The Liouville Eq.~\ref{F4} can
be replaced by a set of coupled equations for the 
mean values of the relevant operators or the Lagrange 
multipliers as follows \cite{ali89}:
 \begin {equation}
\frac{d<\! \hat O_{j} \!>_{t}}{dt} 
= -\sum_{i=0}^{L} g_{ij} <\! \hat O_{i} \!> \; ,
\hspace{1.0cm} \mbox{j = 0, 1, \ldots, L} \; ,\label{F6} 
\end{equation}
\begin {equation}
\frac{d\lambda{j}}{dt} = \sum_{i=0}^{L} \lambda_{i} g_{ji},
\hspace{1.0cm} \mbox{j = 0, 1,\ldots, L}.\label{F7}
\end{equation}
In the MEP formalism, the mean value of the operators 
and the Lagrange multipliers belongs to dual spaces 
which are related by \cite{pro89}
\begin{equation}
<\! \hat O_{j} \!>=-\frac{\partial\lambda_{0} }{ \partial \lambda_{j}}.
\label{F8}
\end{equation}
%%%%%%%%%%%%%%%%%%%%%%%%%%%%%%%%%%%%%%%%%%%%%%%%%%%%%%%%%%%
\section{The Relevant Operators and Evolution Equations}
The {\em generalized time-dependent m-photon JCH} in the RWA 
takes the form
\begin{equation}
\hat H =  E_{1} \hat b_{1}^{\dagger} \hat b_{1} +
E_{2} \hat b_{2}^{\dagger} \hat b_{2} +
\omega \hat a^{\dagger} \hat a   
+  T(t) \Bigl( \gamma \hat
a ^{m} \hat b_{1} \hat b_{2}^{\dagger}
+ \gamma^{*}
\hat b_{2} \hat b_{1}^{\dagger}\hat a^{\dagger m} \Bigr),\label{F9}  
\end{equation}
($\hbar = 1$), where $\gamma$ is the coupling constant between the
system and the external field, $E_{j}$ and $\omega$ are
the energies of the levels and the field, respectively, $\hat a^{\dagger}$,
$\hat a$, are boson operators, $\hat b_{j}^{\dagger}$ and $\hat b_{j}$ are
fermion  operators and $T(t)$ is an arbitrary function of time. 
On setting $m=1$ we recover the {\em generalized time-dependent
JCH} of Gruver et al., \cite{gru93}.

	Working in the context of the generalized time-dependent
JCH Gruver et al.,\cite{gru93} found that the relevant operators 
can be presented in {\it three different but equivalent forms}, 
each of them having different physical interpretations.
These sets are connected via isomorphisms which allows one
to go from one set into another. As noted in \cite{gru93}
the advantage of this multiple representation comes from the fact that
when partial information in any set is known, for instance the initial
values only for some operators are known, it is possible to complete the
missing information via the isomorphisms, if the complementary data in  
any other set is known, i.e. mixed initial conditions \cite{pro89}.
We find that working in the framework of the Hamiltonian \ref{F9}
the same arguments go through. From the form of the Hamiltonian
\ref{F9}, we can guess by looking at the level's population 
and the structure of the interaction terms that a basic set of
relevant operators satisfying Eq.~\ref{F5} is
\begin{eqnarray}
\hat N_{1} & = & \hat b_{1}^{\dagger} \hat b_{1} \; ,\label{F10}\\[0.5ex]
\hat N_{2} & = & \hat b_{2}^{\dagger} \hat b_{2} \; ,\label{F11}\\[0.5ex]
\hat \Delta & = & \hat a^{\dagger} \hat a        \; ,\label{F12}\\[0.5ex]
\hat I^{m} & = &  
\gamma \hat
a^{m} \hat b_{1} \hat b_{2}^{\dagger}
+ \gamma^{*}
\hat b_{2} \hat b_{1}^{\dagger}\hat a^{\dagger m}  \; ,\label{F13}\\[0.5ex]
\hat F^{m} & = & 
i(\gamma \hat   
a ^{m} \hat b_{1} \hat b_{2}^{\dagger}
- \gamma^{*}
\hat b_{2} \hat b_{1}^{\dagger}\hat a^{\dagger m}) \; ,\label{F14}\\[0.5ex]
\hat N_{2,1} & = & \hat b_{2}^{\dagger} \hat b_{2}
		   \hat b_{1}^{\dagger} \hat b_{1} \; .\label{F15}
\end{eqnarray}
These fundamental operators appear in the three possible sets
of relevant operators outlined below. The above operators possess
a simple physical interpretation. $\hat N_{l}$ is the number or
population operator for level one, $\hat N_{2}$ is the number
operator for level two. $\hat \Delta$ is the familiar number
operator for the photon/external field. The operator $\hat{I}^{m}$ 
may be considered as representing the interaction energy between 
the levels and the external the field. The particle's current between 
levels is governed by $\hat{F}^{m}$. $\hat N_{2,1}$ is the 
double occupation number operator.
The operators \ref{F10}-\ref{F11} and \ref{F13}-\ref{F14} can be
considered as the m-photon quantum counter parts of the operators 
obtained for the semiclassical 2-level system studied in 
ref.~\cite{ali91}.

The simplest set which closes a semi-Lie algebra with 
the Hamiltonian, Eq.~\ref{F5}, is found to be,
\begin{eqnarray}
\hat N_{1}^{n} & = &
(\hat a^{\dagger})^{n} \;
\hat N_{1} \;
(\hat a)^{n},\label{F16}\\[0.5ex]
\hat N_{2}^{n} & = &
(\hat a^{\dagger})^{n} \;
\hat N_{2} \; 
(\hat a)^{n},\label{F17}\\[0.5ex]
\hat \Delta^{n} & = &
(\hat a^{\dagger})^{n} \;
\hat \Delta \;
(\hat a)^{n},\label{F18}\\[0.5ex]
\hat I^{n,m} & = &
(\hat a^{\dagger})^{n} \;
\hat I^{m} \;
(\hat a)^{n} \; ,\label{F19}\\[0.5ex]
\hat F^{n,m} & = &
(\hat a^{\dagger})^{n} \;
\hat F^{m} \;
(\hat a )^{n} \; ,\label{F20}\\[0.5ex]
\hat N^{n}_{2,1} & = &
(\hat a^{\dagger})^{n} \;
\hat N_{2,1} \;
(\hat a)^{n} \; ,\label{F21}
\end{eqnarray}
$n = 0, 1, \dots$. 
For $n = 0$ Eqs.~\ref{F16}-\ref{F21} reduce to the fundamental set
of operators given in Eqs.~\ref{F10} through \ref{F15}. This set 
of relevant operators is suitable for numerical simulation as it 
provides the simplest form of the system of differential equations 
for the evolution of their mean values. Eqs.~\ref{F16}-\ref{F21}
are the fundamental operators sandwiched between powers of 
creation $\hat a^{\dagger}$, and destruction $\hat a$ photon 
operators. This leads us to consider the operators 
with $n > 1$ as a measure of virtual transitions due to the 
absorption of more than one photon followed by the emission of 
the extra photons in a transition between the levels.

It is important to emphasize, as pointed out in \cite{gru93}
that the assumption of RWA made at the beginning introduces 
in a natural way the set of correlation functions.
For example if works out the commutator of the particle's 
current between levels, $\hat F^{m}$, and the Hamiltonian
one obtains correlation operators involving the population
operators of levels with the field operators, see Eq.~\ref{F24}
below. Specializing Eq.~\ref{F24} to the simplest case, 
$n=0, m=1$, we have averages of terms such as 
$\hat{N}_{1}^{1}=\hat{a}^{\dagger}\hat{N}_{1}\hat{a}$,
which represent correlation between the field ($\hat{a}$
and $\hat{a}^{\dagger}$) and population of level one 
($\hat{N}_1$). Clearly if one did not impose RWA, the structure 
of the algebra would be very different. Thus the algebra of
the relevant operators carry the very blueprint of the
physical assumptions we make. In short as is evident 
from Eq.~\ref{F5} that the generator of the algebra, the
Hamiltonian, generates a set of operators which are 
closely related to the physics of the problem and in 
this sense we mean that the set is {\em physically} relevant,
thus it is not surprising that any physical assumptions 
that are put into the Hamiltonian will be ``transferred''
to the operators which close the algebra with the Hamiltonian. 

The other two sets of relevant operators which satisfy 
Eq.~\ref{F5} are
\vspace{0.3ex}
\begin{displaymath}
\left\{\frac{1}{2} \left [\hat O_{i} \; (\hat a^{\dagger})^{n} (\hat a)^{n}
+ (\hat a^{\dagger})^{n} (\hat a)^{n} \hat O_{i} \right]\right\}_{n=0}
^{\infty}
\end{displaymath}
and 
\begin{displaymath}
\left\{\frac{1}{2} \left [\hat O_{i} \; (\hat a^{\dagger} \hat a)^{n}
+ (\hat a^{\dagger} \hat a)^{n} \; \hat O_{i}
\right]\right\}_{n=0}^{\infty}.
\end{displaymath} 
Here $\hat O_{i}$ are the fundamental operators
given by Eq.~\ref{F10} through Eq.~\ref{F15}. 
This once again generalizes the $m=1$ case considered
by \cite{gru93}. As pointed out in \cite{gru93} the first 
set can be interpreted as the correlation functions between 
the fundamental operators and $(\hat a^{\dagger})^{n}(\hat a)^{n}$.
$(\hat a^{\dagger})^{n}(\hat a)^{n}$ are proportional to 
the $n$th-order coherence function of the field, \cite{mey90}. 
The operators included in the second set are
proportional to the correlations between the fundamental 
operators and the energy of the field. 

	The dynamical equations for the operators given in
Eq.~\ref{F16}-\ref{F21} can be obtained using
the Ehrenfest theorem (Eq.~\ref{F6}), and are given by
%%%%%%%%%%%%%%%%%%%%%%%%%%%%%%%%%%%%%%%%%%%%%%%%%%%%%%%%%%%%%%%
\begin{eqnarray}
\frac{d<\!\hat N_{1}^{n}\!>}{dt} & = &
T(t)<\!\hat F^{n,m}\!>
+ T(t)[^{n}C_{1}m<\!\hat F^{n-1,m}\!>\nonumber\\ [1.0ex]
& &+^{n}C_{2}m(m-1)<\!\hat F^{n-2,m}\!>+\nonumber\\[1.0ex]
& &....+^{n}C_{m}m(m-1)(m-2).....<\!\hat F^{0,m}\!>]
,\label{F22}\\[1.0ex]
%XXXXXXXXXXXXXXXXXXXXXXXXXXXXXXXXXXXXXXXXXXXXXXXXXXXXXXXXXXXXX
\frac{d<\!\hat N_{2}^{n}\!>}{dt} & = &
-T(t)<\!\hat F^{n,m}\!>,\label{F23}\\[1.0ex]
%XXXXXXXXXXXXXXXXXXXXXXXXXXXXXXXXXXXXXXXXXXXXXXXXXXXXXXXXXXXX
\frac{d<\!\hat F^{n,m}\!>}{dt} & = & - \alpha
<\!\hat I^{n,m}\!>+2|\gamma|^{2}T(t)[
%(n+1)<\!\hat
%N_{2}^{n}\!>\nonumber\\[1.0ex]
-<\!\hat N_{1}^{n+m}\!>\nonumber\\[1.0ex]
& &+<\!\hat N_{2}^{n+m}\!>
+^{n+m}C_{1}m(<\!\hat N_{2}^{n+m-1}\!>
-<\!\hat N_{2,1}^{n+m-1}\!>)\nonumber\\[1.0ex]
& &+^{n+m}C_{2}m(m-1)(<\!\hat N_{2}^{n+m-1}\!>
-<\!\hat N_{2,1}^{n+m-1}\!>)+...],
\label{F24}\\[1.0ex]
%XXXXXXXXXXXXXXXXXXXXXXXXXXXXXXXXXXXXXXXXXXXXXXXXXXXXXXXXXXXX
\frac{d<\!\hat I^{n,m}\!>}{dt} & = & \alpha
<\!\hat F^{n,m}\!>, \label{F25}\\[1.0ex]
%XXXXXXXXXXXXXXXXXXXXXXXXXXXXXXXXXXXXXXXXXXXXXXXXXXXXXXXXXXXX
\frac{d<\!\hat \Delta^{n}\!>}{dt} & = &
+T(t)[^{n+1}C_{1}m<\!\hat F^{n,m}\!>
+^{n+1}C_{2}m(m-1)<\!\hat F^{n-1,m}\!>+....\nonumber \\[1.0ex]
& &+^{n+1}C_{m}m(m-1)(m-2)...<\!\hat F^{0,m}\!>]
%(n+1)T(t)<\!\hat F^{n}\!>
, \label{F26}\\[1.0ex]
%XXXXXXXXXXXXXXXXXXXXXXXXXXXXXXXXXXXXXXXXXXXXXXXXXXXXXXXXXXX
\frac{d<\!\hat  N_{2,1}^{n}\!>}{dt} & = & 0 ,\label{F27}
\end{eqnarray}\\[0.3ex]
%%%%%%%%%%%%%%%%%%%%%%%%%%%%%%%%%%%%%%%%%%%XXXXXXXXXXXXXXXX
$n = 0,1,\dots$, where $\alpha=E_{2}-E_{1}-\omega$
and $^{n}C_{m}=\frac{n!}{(n-m)!m!}$.
Eqs. \ref{F22}-\ref{F27} are the exact dynamical evolution 
equations of the relevant operators for 
the {\em generalized time-dependent m-photon JCH}. 
They can be thought of as a kind of generalized Bloch equations 
for the quantum field case. As can be seen, the different 
order correlations are connected via
the operators $\hat N_{1}^{n}$ and $\hat F^{n,m}$ 
, see Eq.~\ref{F22} and \ref{F24}. 
As a check, if set $m=1$ in Eqs.~\ref{F22} through \ref{F27}
we find that these reduce to the equations resulting 
from the single photon generalized time-dependent JCH
considered by Gruver et al., \cite{gru93}, see their
Eqs.~22-27.

Comparing our fundamental set of operators, viz Eq.~\ref{F10}
through Eq.~\ref{F15} with the one given in \cite{gru93}
we note that the operators which have not changed are
$\hat{N}_1$, $\hat{N}_1$, $\hat{\Delta}_1$ and $\hat{N}_{2,1}$.
Thus one would expect that [although in our case we have
more involved correlations between  
$\hat N_{1}^{n}$ and $\hat F^{n,m}$],   
\begin{equation}
\left\{<(\hat a^{\dagger})^{n} \;   
\hat N_{1} \;
(\hat a)^{n}> +
<(\hat a^{\dagger})^{n} \;
\hat N_{2} \;
(\hat a)^{n}> - <
(\hat a^{\dagger})^{n-1}\Delta(\hat a)^{n-1}>\right\}^{\infty}_{n=0} ,
\label{F28} 
\end{equation}
and
%%%%%%%%%%%%%%%%%%%%%%%%%%%%%%%%%%%%%%%%%%%%%%%%%
\begin{equation}
\left\{<(\hat a^{\dagger})^{n}
\hat N_{2,1}
(\hat a)^{n}>\right\}^{\infty}_{n=0} ,\label{F29}
\end{equation}\\[0.3ex]
are constants of the motion. Indeed we find this to be the case
as can be seen from Eqs.~\ref{F22}-\ref{F27}. It is easy
to see that \ref{F29} holds by directly looking at Eq.~\ref{F27}.
To check the validity of \ref{F28} we first rewrite Eq.~\ref{F26}
for $n-1$
\begin{eqnarray}
\frac{d<\!\hat \Delta^{n-1}\!>}{dt} & = &
+T(t)[^{n}C_{1}m<\!\hat F^{n-1,m}\!>
+^{n}C_{2}m(m-1)<\!\hat F^{n-2,m}\!>+....\nonumber\\
& &+^{n}C_{m}m(m-1)(m-2)...<\!\hat F^{0,m}\!>] 
\label{F30}
%\\[1.0ex]
%(n+1)T(t)<\!\hat F^{n}\!>, \label{F30}\\[1.0ex]
\end{eqnarray}
Adding Eqs.~\ref{F22} and \ref{F23} and subtracting \ref{F30}
we immediately see that \ref{F28} holds. This provides an
excellent check on our manipulations which are tedious.
From the expressions \ref{F28} and \ref{F29} it follow
that the particle's current between levels is equal to the 
photon's flux. In the case, $n=0$ we end-up with the 
conservation of the level's population. For $n>0$ we obtain 
a restriction for the correlations.

	It is clear from Eq.~\ref{F28} that the mean 
value of the operators are not be independent. This 
restricts the choice of the initial conditions. It is 
therefore necessary to choose a formalism which respects 
the restriction on the initial conditions. Following 
ref.~\cite{gru93} one may settle this issue by using the 
MEP density matrix given in Eq.~\ref{F1}. As is known in 
the MEP formalism the mean values and the Lagrange multipliers 
live in the dual spaces. As mentioned in \cite{gru93} the 
Lagrange multipliers are numbers that can be freely chosen.
Now once the restriction Eq.~\ref{F28} on the mean values
is implemented the equivalent restriction for the 
Lagrange multipliers are automatically satisfied when
the density operator is diagonalized. In the MEP
formalism used in \cite{gru93} lack of knowledge on the 
mean value of one operator is equivalent to setting its 
Lagrange multiplier equal to zero.
%%%%%%%%%%%%%%%%%%%%%%%%%%%%%%%%%%%%%%%%%%%%%%%%%%%%%%%%%
\section{Conclusions}
We have presented a generalized version of the 
m-photon {\em JCH} giving a description in terms 
of physical relevant operators. The temporal evolution 
equations have been worked out for one of the set
which provides the simplest form of the system
of differential equations for the evolution
of the mean values of the operators. 
Since an arbitrary function of time has been included, 
this formalism allows us to study the system even when 
the coupling is time dependent. 
Our work is a simple extension of the work of 
Gruver et al.~\cite{gru93} to the m-photon case. 
One advantage, as mentioned in \cite{gru93}, 
of giving a description of the system in terms of 
the three sets of physical relevant 
operators is that it allows 
one to treat the case of mixed boundary conditions.
%%%%%%%%%%%%%%%%%%%%%%%%%%%%%%%%%%%%%%%%%%%%%%%%%%%%%%%%%%%%
\section*{Acknowledgments}
We thank J.~L.~Gruver for suggesting this problem.

\end{document}